\documentclass[alpha-refs]{wiley-article}

\usepackage{color}
\usepackage{ulem}

\usepackage{siunitx}
\usepackage[pdfencoding=auto]{hyperref}
\usepackage{subcaption}

\papertype{Research Article}
\paperfield{Quarterly Journal of the Royal Meterological Society}

\title{Multistability in a Coupled Ocean-Atmosphere Reduced Order Model: Non-linear Temperature Equations}

\abbrevs{LFV, Low Frequency Variability; AMOC, Atlantic Meridional Overturning Circulation; NAO, North Atlantic Osciliation; MAOOAM, Modular Arbitrarty Order Ocean Atmosphere Model.}

\author[1, 2]{Oisín Hamilton}
\author[1]{Jonathan Demaeyer}
\author[1]{Stéphane Vannitsem}
\author[2]{Michel Crucifix}

\affil[1]{Climate Dynamics, Royal Meterological Institute of Belgium, Brussels, Belgium}
\affil[2]{Earth and Life Institute, Université catholique de Louvain, Louvain-la-Neuve, Belgium}

\corraddress{Oisín Hamilton, Climate Dynamics, Royal Meterological Institute of Belgium, Uccle, Brussels, 1180, Belgium}
\corremail{oisin.hamilton@meteo.be}

\fundinginfo{European Union Horizon 2020, Marie Sklodowska-Curie grant agreement No.956170}

\runningauthor{Hamilton et al.}

\begin{document}

\maketitle

\begin{abstract}
    Multistabilities were found in the ocean-atmosphere flow, in a reduced order ocean-atmosphere coupled model, when the non-linear temperature equations were solved numerically. In this paper we explain how the full non-linear Stefan-Bolzmann law was numerically implemented, and the resulting change to the system dynamics compared to the original model where these terms were linearised. Multiple stable solutions were found that display distinct ocean-atmosphere flows, as well as different Lyapunov stability properties. In addition, distinct Low Frequency Variability (LFV) behaviour was observed in stable attractors. We investigated the impact on these solutions of changing the magnitude of the ocean-atmospheric coupling, as well as the atmospheric emissivity to simulate an increasing green-house effect. Where multistabilities exist for fixed parameters, the possibility for tipping between solutions was investigated, but tipping did not occur in this version of the model where there is a constant solar forcing. This study was undertaken using a reduced-order quasi-geostrophic ocean-atmosphere model, consisting of two atmosphere layers, and one ocean layer, implemented in the Python programming language.

    \keywords{Low Frequency Variability, Temperature multistabilities, Coupled ocean-atmosphere model}
\end{abstract}

\section{Introduction}\label{sec:introduction}
    The average temperature of the Earth is increasing as additional longwave radiation is being captured by the atmosphere. Due to non-linear processes in the climate system, it is expected that this global temperature increase could lead to tipping points~\citep{lenton2008}, where certain elements of the climate change rapidly and where some of these changes could be irreversible. Due to the potential ramifications of such changes, understanding what tipping elements are present in the climate, and how close we are to these tipping points is of major importance. 

    In the North Atlantic several tipping elements have been identified, such as the Atlantic Meridional Overturing Circulation (AMOC), Greenland ice sheets, the subpolar gyre, and the North Atlantic jet stream~\citep{steffen2018, armstrongmckay2022}. A change to any of these climate elements would have a great impact on the climate of Europe, in addition, existing long term climate patterns could be fundamentally altered. Low Frequency Variability (hereafter, LFV) in the midlatitudes is a catch-all term that is used to refer to climate processes that slowly vary on interannual to multidecadal time scales. An example of LFV is the North Atlantic Oscillation (NAO), which is the sea-surface pressure difference between the subtropical and subpolar gyres. It is thought that better prediction of LFV could help extend the typical two-week weather forecast horizon, which is caused by the sensitivity to initial conditions, in the midlatitudes.
    
    Potential mechanisms for LFV in the midlatitudes include coupling of the ocean with the atmosphere where temperature, wind stresses, and moisture are transferred from one to another~\citep{holton2004}. This coupling has been observed in data~\citep{arthun2021, wu2005, czaja2002, czaja2001}. Studies based on numerical simulation have attempted to reproduce this LFV using models of varying complexity. One line of research is to use the theory of dynamical systems to isolate the dynamical mechanisms at the origin of the LFV, which is done by investigating models with a low number of degrees of freedom. These models purposely sacrifice the required complexity to accurately model the real climate and they instead focus on the minimum set of conditions necessary to generate the observed variability. 
    
    The approach of using reduced order models in climate science goes back to the work of Lorenz~\citep{lorenz1960}, and more recently coupled ocean-atmosphere models have been developed to investigate the behaviour stemming from coupling the ocean and atmosphere systems~\citep{lorenz1984, roebber1995, vannitsem2014, decruz2016}. The model that we use in this study has been developed by coupling the reduced order atmosphere model developed by Charney and Devore~\citep{charney1979}\footnote{Further developed by Charney and Straus~\citep{charney1980} and again by Reinhold and Pierrehumbert~\citep{reinhold1982}.} with the ocean model from Pierini~\citep{pierini2011}. This model was originally described in~\citet{vannitsem2014}, and later a heat balance was introduced using an energy balance model~\citep{vannitsem2015}. It was the addition of the heat transfers in this latest model version, in conjunction with the wind stress between the ocean and the atmosphere, that produced LFV. This model has since been implemented in the Python programming language, and named the Modular Arbitrary Order Ocean-Atmosphere Model (MAOOAM)~\citep{decruz2016, demaeyer2020}. It is this version that we have used in this study.

    One source of non-linearity in the model equations comes from the long wave radiation emitted from the ocean and atmosphere, and modelled using the Stefan-Boltzmann law ($\sigma_{\text{B}} T^4$, where $\sigma_{\text{B}}$ is the Stefan Bolzmann constant). In the MAOOAM model, the quartic radiation terms are linearised to simplify the projection of the equations onto a truncated Fourier expansion. This linearisation is justified by the fact that the perturbations in temperature are small in relation to the climatological reference temperatures~\citep{decruz2016}. However, this linearisation removes the possibilities of non-linear interactions from the temperature terms. For this reason, this study investigates the impact on the LFV in the model, and the potential for multi-stabilities or bifurcations, when the temperature equations are not linearised. 

    We will show that removing the linearisation leads to multiple stable flow patterns in the atmosphere and ocean, for certain levels of ocean-atmosphere coupling and atmosphere emissivity. These flow patterns are qualitatively distinct and result in multiple average temperatures, for the same model parameters. They also present different cycle lengths and different dominant modes.

    Section \ref{sec:Model} describes the reduced order model used in this study, and Section \ref{sec:Model_sub:model_modifications} describes the modifications made to remove the requirement of linearising the temperature equations. This section also gives a description of the model configurations used in this study. The results are split into two sections, where we first look at the results from altering the ocean-atmosphere coupling (Section \ref{subsec:results_coupling}), and then the impact of atmospheric emissivity (Section \ref{subsec:results_emissivity}). In each of the two results sections, we look at the impact of altering the given parameters on the stability and predictability of the system. Section \ref{sec:discussion} summarises the main results and discusses the general implications of these findings.

\section{qgs Model}\label{sec:Model}
\subsection{Model Description}
    The qgs model~\citep{demaeyer2020} is a reduced-order midlatitude climate model, with many different model configurations available. In the present work, we use the ocean-atmosphere model version where the atmospheric flow is obtained from a two-layer quasi-geostrophic flow defined on a $\beta$ plane~\citep{reinhold1982}. Similarly, the ocean streamfunctions are modelled using a quasi-geostrophic shallow-water model with a rigid lid~\citep{pierini2011}. The thermodynamic equation for the atmosphere and ocean temperatures are derived using an energy balance scheme proposed by~\citet{barsugli1998}. The coupled ocean-atmosphere scheme used here was first introduced by Vannitsem et al.~\citep{vannitsem2015}. The atmosphercic variables are coupled through wind stress to the oceanic ones, driving the ocean circulation, which transports heat in the ocean. The ocean transfers heat with the atmosphere through radiative and direct heat coupling, which in turn impacts the atmospheric flow. 

    In this study we imposed a closed ocean basis (no-flux boundary conditions on all boundaries), and a channel atmosphere (no-flux boundary conditions on the north and south boundaries, and periodic boundary conditions at the west and east). We describe how these boundary conditions are implemented in Section \ref{sec:Model_sub:num-sols}. This version of the model with a closed ocean basin coupled to an atmosphere is called the Modular Arbitrary Order Ocean Atmosphere Model (MAOOAM)~\citep{decruz2016}.  

    The governing partial differential equations (PDEs) for the atmosphere barotropic $\psi_{\mathrm{a}}$ and baroclinic $\theta_{\mathrm{a}}$ streamfunctions and ocean streamfunctions $\psi_{\mathrm{o}}$ are given as: 

    \begin{align}
        \frac{\partial}{\partial t}\left(\nabla^2 \psi_{\mathrm{a}}\right)+J\left(\psi_{\mathrm{a}}, \nabla^2 \psi_{\mathrm{a}}\right)+J\left(\theta_{\mathrm{a}}, \nabla^2 \theta_{\mathrm{a}}\right)+\beta \frac{\partial \psi_{\mathrm{a}}}{\partial x} &=-\frac{k_d}{2} \nabla^2\left(\psi_{\mathrm{a}}-\theta_{\mathrm{a}}-\psi_{\mathrm{o}}\right)\\
        \frac{\partial}{\partial t}\left(\nabla^2 \theta_{\mathrm{a}}\right)+J\left(\psi_{\mathrm{a}}, \nabla^2 \theta_{\mathrm{a}}\right)+J\left(\theta_{\mathrm{a}}, \nabla^2 \psi_{\mathrm{a}}\right)+\beta \frac{\partial \theta_{\mathrm{a}}}{\partial x} &=-2 k_d^{\prime} \nabla^2 \theta_{\mathrm{a}}\nonumber \\ 
        &\hspace*{1cm} +\frac{k_d}{2} \nabla^2\left(\psi_{\mathrm{a}}-\theta_{\mathrm{a}}-\psi_{\mathrm{o}}\right)+\frac{f_0}{\Delta p} \omega\\
        \frac{\partial}{\partial t}\left(\nabla^2 \psi_{\mathrm{o}}-\frac{\psi_{\mathrm{o}}}{L_{\mathrm{R}}^2}\right)+J\left(\psi_{\mathrm{o}}, \nabla^2 \psi_{\mathrm{o}}\right)+\beta \frac{\partial \psi_{\mathrm{o}}}{\partial x} &=-r \nabla^2 \psi_{\mathrm{o}}+d \nabla^2\left(\psi_{\mathrm{a}}-\theta_{\mathrm{a}}-\psi_{\mathrm{o}}\right), 
    \end{align}

    where $\psi_{\mathrm{a}}$ and $\psi_{\mathrm{o}}$ are the atmosphere and ocean barotropic streamfunctions, and $\theta_{\mathrm{a}}$ are the atmosphere baroclinic streamfunctions. Vertical velocities are given by $\omega$.

    The ocean and atmosphere temperatures are derived from an energy balance model:
    \begin{align}\label{eq:temp_equations_atm}
        \gamma_{\mathrm{a}}\left(\frac{\partial T_{\mathrm{a}}}{\partial t}+J\left(\psi_{\mathrm{a}}, T_{\mathrm{a}}\right)-\sigma\omega\frac{p}{R}\right)&=-\lambda\left(T_{\mathrm{a}}-T_{\mathrm{o}}\right)+\varepsilon\sigma_{\mathrm{B}} T_{\mathrm{o}}^4-\varepsilon \sigma_{\mathrm{B}} T_{\mathrm{a}}^4+R_{\mathrm{a}}\\
        \label{eq:temp_equations_ocn}
        \gamma_{\mathrm{o}}\left(\frac{\partial T_{\mathrm{o}}}{\partial t}+J\left(\psi_{\mathrm{o}}, T_{\mathrm{o}}\right)\right)&=-\lambda\left(T_{\mathrm{o}}-T_{\mathrm{a}}\right)-\sigma_{\mathrm{B}} T_{\mathrm{o}}^4+\varepsilon \sigma_{\mathrm{B}} T_{\mathrm{a}}^4+R_{\mathrm{o}}, 
    \end{align}

    \begin{table}
        \caption{MAOOAM Model Parameters}
        \centering
        \def\arraystretch{1.5}
        \begin{tabular}[]{|c r p{11cm}|}
            \hline
            \multicolumn{1}{|c}{Parameter} & \multicolumn{1}{c}{Value} & \multicolumn{1}{l|}{Description (units)}\\
            \hline
            $\beta$ & $1.62\times10^{-11}$ & The meridional gradient of the Coriolis parameter at a given latitude ($\si{\metre^{-1}\second^{-1}}$)\\
            $f_0$ & $1.032\times 10^4$ & Coriolis parameter ($\si{\second^{-1}}$)\\
            $k_d$ & $\mathrm{g}C/\Delta p$ & Atmosphere-surface friction ($\si{\second^{-1}}$)\\
            $k_d'$ & $\mathrm{g}C/\Delta p$ & Internal atmosphere friction ($\si{\second^{-1}}$)\\
            $r$ & $ 1\times 10^{-7} $ & Ocean bottom Rayleigh friction ($\si{\second^{-1}}$)\\
            $L_R$ & $1.9934\times10^4$ & Reduced Rossby deformation radius\\
            $h$ & $136.5$ & Depth of the ocean layer ($\si{\metre}$)\\
            $d$ & $C/(\rho_{\mathrm{o}}h)$ & Coefficient of mechanical ocean-atmosphere coupling ($\si{\second^{-1}}$)\\
            $\gamma_{\mathrm{a}}$ & $1\times10^7$ & Specific heat capacity of the atmosphere ($\si{\joule\metre^{-2}\kelvin}$)\\
            $\gamma_{\mathrm{o}}$ & $5.46\times10^8$ & Specific heat capacity of the ocean ($\si{\joule\metre^{-2}\kelvin^{-1}}$)\\
            $\sigma$ & $0.2$ & Static stability of the atmosphere\\
            $\lambda$ & $1004C$ & Sensible and turbulent heat exchange between ocean and atmosphere ($\si{\watt\metre^{-2}\kelvin^{-1}}$)\\
            $R$ & $287.058$ & Gas constant in dry air ($\si{\joule\kilogram^{-1}\kelvin^{-1}}$)\\
            $\sigma_{\mathrm{B}}$ & $5.67\times 10^{-8}$ & Stefan-Boltzmann constant ($\si{\joule\metre^{-2}\second^{-1}\kelvin^{-4}}$)\\
            \hline
        \end{tabular}
        \caption*{The key model parameters used in this study. Here $C$ is a variable that we alter to investigate the impact of the ocean-atmosphere coupling. As in~\citet{charney1980} we assume $k'=k_d$ and as in~\citet{vannitsem2015a} that gravity $\mathrm{g}=10\si{\metre\second^{-2}}$.}
        \label{tab:model_parameters}
    \end{table}

    where $T_{\mathrm{a}}$ and $T_{\mathrm{o}}$ are the atmosphere and ocean temperatures, $\gamma_{\mathrm{a}}$ and $\gamma_{\mathrm{o}}$ are the heat capacities of the atmosphere and ocean, $\sigma$ is the static stability of the atmosphere (assumed constant), $\sigma_{\mathrm{B}}$ is the Stefan-Boltzmann constant, $R_{\mathrm{a}}$ and $R_{\mathrm{o}}$ are the incoming solar radiation absorbed by the atmosphere and ocean, and $\varepsilon$ is the atmospheric emissivity.

    To reduce the number of variables, the atmosphere temperature variable $T_{\mathrm{a}}$ is related to the baroclinic streamfunctions using the hydrostatic balance in pressure coordinates and the ideal gas law, providing the relationship $T_{\mathrm{a}}=2f_0\theta_{\mathrm{a}}/R$.

\subsection{Numerical Solution}\label{sec:Model_sub:num-sols}
    The differential equations are projected onto basis modes, a procedure also known as \emph{Galerkin expansion}. The basis modes are chosen to ensure that the boundary conditions described in the previous section are satisfied. This is done by stipulating that $\phi_i(x,y)=0$ for points $(x, y)$ on the boundary, and $\frac{\partial F_i(x, y)}{\partial x}=0$ for $x$ on the boundary and $F_i(x, y)=0$ for $y$ on the boundary. In this study we use 10 basis modes for the atmosphere and 8 for the ocean, as in~\citet{vannitsem2017}. These are set on a domain of $x\in[0, 2\pi],\ y\in[0, \pi]$. The atmosphere $F_i$ and ocean $\phi_i$ modes are given below:

    \begin{equation}
        \begin{aligned}[c]
            F_1&=\sqrt{2} \cos(y)\\
            F_2&=2\cos(nx)\sin(y)\\
            F_3&=2\sin(nx)\sin(y)\\
            F_4&=\sqrt{2}\cos(2y)\\
            F_5&=2\cos(nx)\sin(2y)\\
            F_6&=2\sin(nx)\sin(2y)\\
            F_7&=2\cos(2nx)\sin(y)\\
            F_8&=2\sin(2nx)\sin(y)\\
            F_9&=2\cos(2nx)\sin(2y)\\
            F_{10}&=2\sin(2nx)\sin(2y)\\
        \end{aligned}
        \qquad \qquad
        \begin{aligned}[c]
            \phi_1&=2\sin(nx/2)\sin(y)\\
            \phi_2&=2\sin(nx/2)\sin(2y)\\
            \phi_3&=2\sin(nx/2)\sin(3y)\\
            \phi_4&=2\sin(nx/2)\sin(4y)\\
            \phi_5&=2\sin(nx)\sin(y)\\
            \phi_6&=2\sin(nx)\sin(2y)\\
            \phi_7&=2\sin(nx)\sin(3y)\\
            \phi_8&=2\sin(nx)\sin(4y)\\
        \end{aligned}
    \end{equation}
    
    Three basis modes are of particular interest as they have real world analogies: 
    \begin{itemize}
        \item $F_1(x, y)=\sqrt{2}\cos(y)$ represents the solar insolation imbalance between the north and south.
        \item $\phi_1(x, y)=2\sin(x/2)\sin(y)$ represents average temperature fluctuations in the ocean.
        \item $\phi_2(x, y) =2\sin(x/2)\sin(2y)$ is the double gyre, orientated so the peak is either to the north or south of the tough. This loosely approximates the NAO, which is defined by the difference in surface pressure anomalies between northern and southern locations (often the Azores and Iceland)~\citep{hurrell2003}. The prevailing clockwise winds around the Azores high, and the counter clockwise winds around the northern low pressure can be broadly simulated by projecting the atmospheric streamfunction anomalies on this mode, thus simulating the impact on the wind and heat transport caused by the NAO.
    \end{itemize}\

    The model variables are expanded using the basis modes. In previous studies using such energy balance models the temperature variables in the model are linearised around a fixed in time equilibria temperature: $T(t, x, y)=T_0+\delta T(t, x, y)$ to remove the quartic terms $\sigma_{\mathrm{B}}T^4$~\citep{vannitsem2015}. This resulted in the temperatures being expressed as:
    
    \begin{equation}\label{eq:temperature_linearised_expansion}
        \begin{aligned}
            T_{\mathrm{a}}(t,x,y) &= T_{\mathrm{a}, 0} + \delta T_{\mathrm{a}}(t, x, y) &= T_{\mathrm{a}, 0} + \sum_{i=1}^{10}\delta T_{\mathrm{a}, i}(t)F_{i}(x, y)\\ 
            T_{o}(t,x,y) &= T_{o, 0} + \delta T_{o}(t, x, y) &= T_{o, 0} + \sum_{i=1}^{8}\delta T_{o, i}(t)\phi_{i}(x, y). 
        \end{aligned}
    \end{equation}

    The PDEs introduced in Equations (\ref{eq:temp_equations_atm}, \ref{eq:temp_equations_ocn}) are then projected onto these basis modes, using the inner product:

    \begin{equation}
        \langle f, g\rangle = \frac{n}{2\pi^2}\int_0^\pi\int_0^{2\pi/n}f(x, y)g(x, y)\mathrm{d}x \mathrm{d}y
    \end{equation}

    This leads to 20 ordinary differential equations (ODEs) for the atmospheric streamfunctions, 10 for the barotropic and 10 for the baroclinic streamfunctions. In addition there are 16 ODEs in the ocean, 8 for the barotropic streamfunctions and 8 for the temperature anomaly. This leads to a total of 36 ODEs describing the model.

\subsection{Non-linear radiation terms}\label{sec:Model_sub:model_modifications}
    This study focuses on the change in the system dynamics caused by not linearising the radiation terms in the temperature equations of the MAOOAM model. This requires the reference temperature $T_{\mathrm{a}, 0}$ and $T_{\mathrm{o}, 0}$ to be time-varying quantities in the expansions shown in Equation~\ref{eq:temperature_linearised_expansion}. Therefore, to allow the average temperature across the atmosphere and ocean to change dynamically with time, we introduced two new basis modes, corresponding to constant spatial modes: $F_0(x, y)=1$ and $\phi_0(x, y)=1$. These modes were added to the list of basis modes that were introduced in Section \ref{sec:Model_sub:num-sols}:

    \begin{equation}
        \begin{aligned}[c]
            F_0&=1\\
            F_1&=\sqrt{2} \cos(y)\\
            F_2&=2\cos(nx)\sin(y)\\
            &\ \ \vdots
        \end{aligned}
        \qquad \qquad
        \begin{aligned}[c]
            \phi_0&=1\\
            \phi_1&=2\sin(nx/2)\sin(y)\\
            \phi_2&=2\sin(nx/2)\sin(2y)\\
            &\ \ \vdots
        \end{aligned}
    \end{equation}

    These additional basis modes now allow the expansions, shown in Equation \ref{eq:temperature_linearised_expansion}, to be given as $T_{o}(t,x,y)=\sum_{i=0}^{8}T_{o, i}(t)\phi(x, y)$, and similarly for the atmosphere.

    Only the temperature equations are projected onto these additional basis modes, as these are the variables linearised in the original MAOOAM model. This means that we still have 10 ODEs for the atmospheric barotropic streamfunctions, and 8 ODEs for the oceanic barotropic streamfunctions. We obtain an additional ODE for the atmospheric baroclinic streamfunction (as this variable replaces the atmospheric temperature variables) and for the ocean temperature. This increases the total number of ODEs describing the system to 38. 
    
    We now introduce the two modified versions of the model that are used in the current study:  

    \textbf{Dynamic Equilibria}: this version of the model includes the same linearisation as the Linearised version of the model, but the equilibrium temperature is dependant on time: $T(t, x, y)=T_0(t)+\delta T(t, x, y)$. We will refer to this version as DE.

    \textbf{Non-Linear T4}: This version does not involve the linearisation of the Stefan-Boltzmann law terms and the equations are projected directly onto the basis modes. This retains the quartic radiation terms. We will refer to this version of the model as T4.

    This results in three model versions:
    \begin{itemize}
        \item Linear Model (LM)
        \item Dynamic Equilibria (DE)
        \item Non-Linear (T4)
    \end{itemize}

    More information about how these modifications were made can be found in the \href{https://qgs.readthedocs.io/en/latest/}{model manual}~\citep{qgs_model}. See also the Supporting Information.
\section{Results}\label{sec:results}
    The results section is split into two main parts. We first describe the system dynamics when the ocean atmosphere coupling is altered, and second we alter the atmospheric emissivity. We focus on values of these parameters that result in multiple distinct ocean-atmosphere flows, which are described by distinct attractors. Multiple stable attractors, for given parameter values, provide the possibility for model solutions to transition from one attractor to another, provided appropriate forcing is imposed. Such transitions between attractors could represent tipping points or abrupt changes or switching in flows. We are particularly interested in flows that present LFV, generated by the coupling between ocean and atmosphere variables, due to the increased predictability of the system dynamics that these solutions provide.

    \begin{figure}
        \centering
        \includegraphics[width=\textwidth]{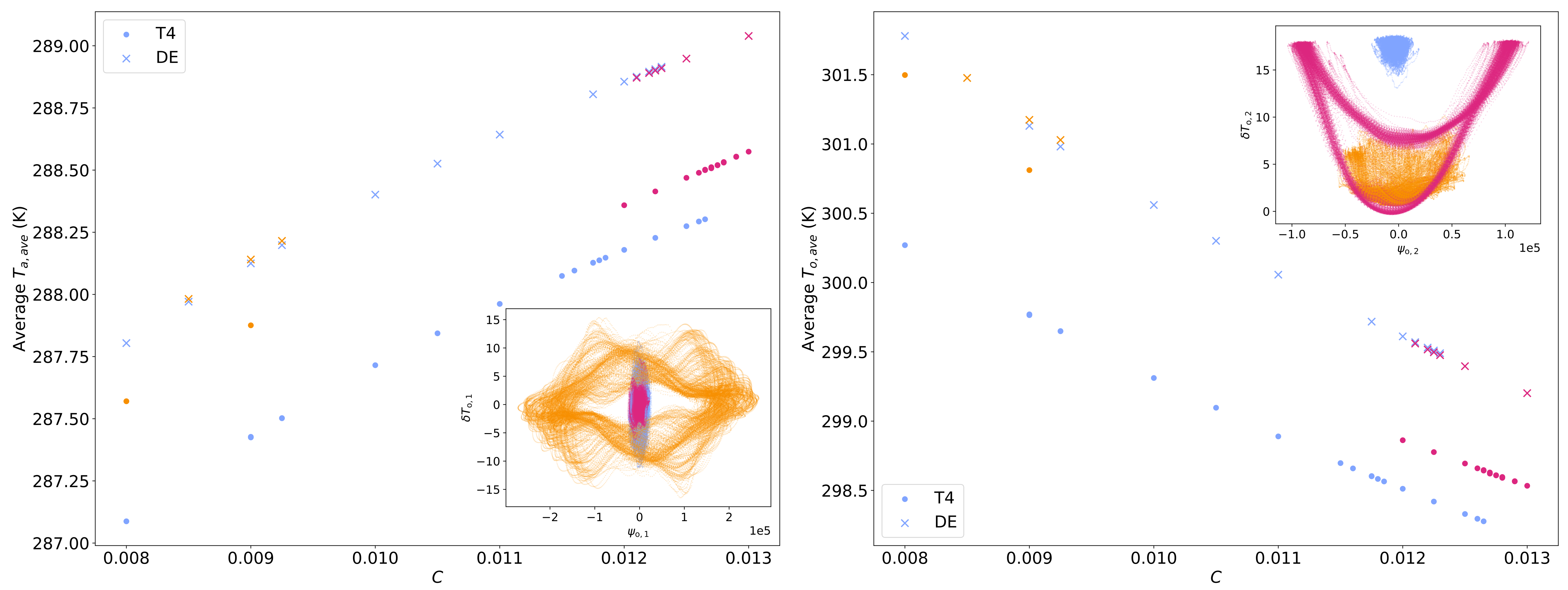}
        \caption{The Temperature-C diagrams for the averaged atmosphere (left), and ocean (right) temperatures for the DE model (crosses), and the T4 mode (dots), of the stable attractors. The branches were colour coded by qualitatively investigating the behaviour of each attractor. The attractors are projected onto the planes $(\psi_{o, 1}, T_{o, 1})$ (left) and $(\psi_{o, 2}, T_{o, 2})$ (right), to display how the behaviour differs between the three stable attractors. The projection of the orange attractor is shown for $C=0.009\si{\kilogram\metre^{-2}\second^{-1}}$ and the other two projections (blue and pink) of the attractors are shown for $C=0.0125\si{\kilogram\metre^{-2}\second^{-1}}$.}
        \label{fig:temp_bif_diag_C}
    \end{figure}

\subsection{Ocean-Atmosphere Coupling}\label{subsec:results_coupling}
    This section presents the results where the magnitude of ocean-atmosphere coupling is varied for the three model versions. Increasing the ocean-atmosphere coupling has the effect of increasing the heat transfer between the ocean and atmosphere. This reduces the ocean temperature (which has a higher equilibria temperature than the atmosphere) and increases the atmospheric temperature, which has an impact on the baroclinic streamfunctions. The ocean temperature anomalies cause uneven temperature exchanges in the atmosphere that require the heat energy to be transported by atmospheric winds. At the same time, increasing the ocean-atmosphere coupling increases the friction felt by the atmosphere from the surface wind stresses with the ocean. In turn, this causes the atmospheric wind to have a greater impact on the movement of the sea temperature anomalies. Together, this reduces the fast moving timescales of the atmosphere through coupling to the ocean, which has a slower timescale relative to the atmosphere.
    
    We alter the ocean-atmosphere coupling $C$ by altering the following parameters: the strength of the ocean-atmosphere coupling $d$, the ocean-atmosphere friction $k_d$, the internal atmosphere friction $k_d'$, and the direct heat transfer between the ocean and atmosphere $\lambda$~\citep{vannitsem2015a}. These parameters are controlled using a single friction coefficient $C$, where the relationship between $C$ and the above parameters is given in Table \ref{tab:model_parameters}. In this study we focus on values of $C$ that are deemed to be within a realistic range ($C\in [0.008\si{\kilogram\metre^{-2}\second^{-1}},\ 0.02\si{\kilogram\metre^{-2}\second^{-1}}]$) for the real world coupling of the ocean and atmosphere~\citep{houghton1986, nese1993, vannitsem2017}.

\subsubsection{State Space Probing}
    \begin{figure}
        \centering
        \includegraphics[width=\textwidth]{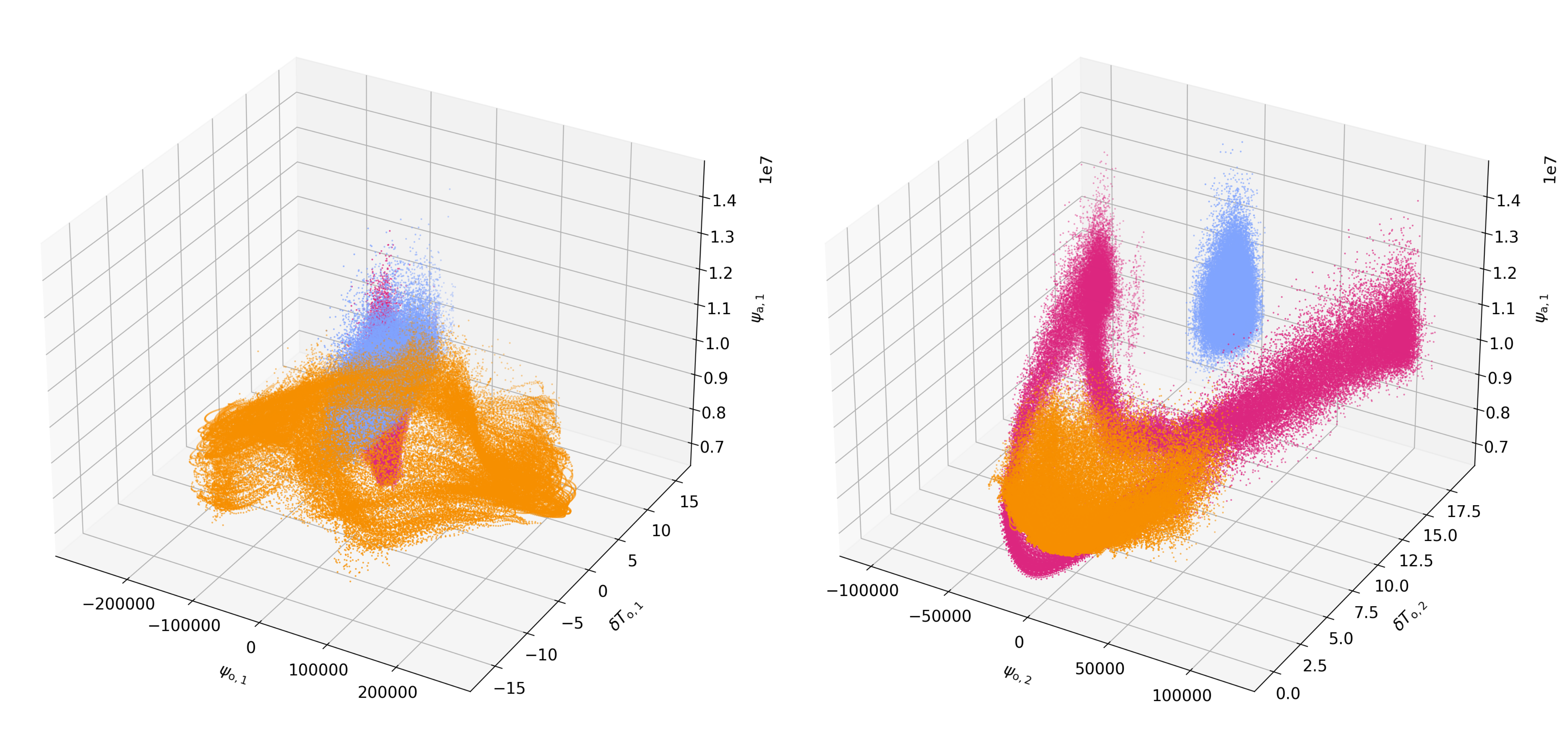}
        \caption{The distinct attractors for the T4 runs are projected onto $(\psi_{\mathrm{a}, 1}, \psi_{o, 1}, T_{o, 1})$ (left) and $(\psi_{\mathrm{a}, 1}, \psi_{o, 2}, T_{o, 2})$ (right), to present the level of ocean-atmosphere coupling. The orange attractor is shown for $C=0.009\si{\kilogram\metre^{-2}\second^{-1}}$ and the other two attractors are shown for $C=0.0125\si{\kilogram\metre^{-2}\second^{-1}}$. Here we only present the T4 runs as the dynamics over these three variables are identical between the T4 and DE model runs.}
        \label{fig:3d_ocn_atm_C}
    \end{figure}

    The effect of the ocean-atmosphere coupling on the system dynamics was investigated by fixing the coupling value $C$ and running many trajectories with random initial conditions. Once these trajectories had appeared to settle onto an attractor, the initial transient section of the trajectories was discarded and we continued to run the trajectories for long run times ($1\times 10^7$ model days). This was done to ensure that the attractors remain stable and trajectories remain on the attractor. For these experiments, the atmosphere emissivity was set to $\varepsilon=0.7$ as the model default~\citep{decruz2016}. This process was repeated for different values of $C$.

    The average temperatures of the ocean and atmosphere were calculated by projecting the trajectories, that were embedded within an attractor, onto the basis modes to obtain the temperature profiles for each time step. We then took the average across the spatial domain to obtain a single average temperature for each time step. Finally we took the average across time to obtain a single temperature, which represents the average temperature associated with a given attractor. This process is shown for the temperature of the ocean, with a similar process taken to calculate the average atmospheric temperatures:

    \begin{equation*}
        \langle T_o(t)\rangle_{\text{space, time}} = \frac{1}{n_\tau n_xn_y}\sum_{\tau=0}^{n_\tau}\sum_{i=0}^{n_x}\sum_{j=0}^{n_y}\sum_{k=0}^{n_o} T_{o, k}(t_\tau)\phi_k(x_i, y_j)
    \end{equation*}

    Where $n_o$ are the number of ocean modes, $n_x$ and $n_y$ are the number of spatial grid points being averaged across, and $n_\tau$ is the number of time steps in the numerical solution.

\subsubsection{Multistabilies in Temperature - \texorpdfstring{$C$}{}}\label{subsec:temp_multistabilities_c}
    The average spatial and temporal temperatures are presented on a  Temperature-C diagram. These diagrams are not bifurcation diagrams, as we only have information about the stable branches that we could find using the described method. The resulting figures for the average atmosphere and ocean temperatures, where the value of the ocean-atmosphere coupling is altered, are shown in Figure \ref{fig:temp_bif_diag_C}. In these figures, the branches are colour coded depending on the qualitative behaviour of the attractor. The figures show there are two intervals of $C$ where there are multistabilies in the temperature. These intervals are approximately $C\in[0.008\si{\kilogram\metre^{-2}\second^{-1}}, 0.009\si{\kilogram\metre^{-2}\second^{-1}}]$ and $C\in[0.012\si{\kilogram\metre^{-2}\second^{-1}}, 0.0127\si{\kilogram\metre^{-2}\second^{-1}}]$. These figures also show that there exists a single stable branch that bridges the two intervals (shown in blue). We plot the results of the dynamic equilibria (DE) and non-linear (T4) model runs on the same plot. 
    
    We found two differences between the T4 and DE runs.  The first is that the difference in average temperatures between the stable branches in the DE runs are smaller than in the T4 runs. The small temperature differences between the attractors in the DE runs occurs due to the zeroth order temperature equations controlling the equilibria temperature $(T_{\mathrm{a}, 0},\ T_{o, 0})$ having only one real stable solution. This means that the difference in average temperatures in each stable equilibrium is caused by only higher order terms. In the T4 model, there are additional terms in the zeroth order temperature equations that result in larger differences between the zeroth order temperatures. This difference in the equations occurs in the non-linear longwave radiation terms, and is sketched below for the ocean temperature equation. In the below expressions we introduce the tensor $v_{i,j,k,l,m}=\langle F_i,\, \phi_j,\, \phi_k,\, \phi_l,\, \phi_m \rangle$.

    \vspace*{0.3cm}
    \begin{tabular}{l p{6cm} p{6cm}}
        & \multicolumn{1}{l}{Full Term}& \multicolumn{1}{l}{Zero Order Term}\\ \\
        DE Model & $T^4_{o, 0}+4T_{o, 0}^3\sum_{m=1}^{n_o}v_{i,0,0,0,m} T_{o, m}$ & $T_{o, 0}^4$\\ \\
        T4 Model & $\sum_{j,k,l,m=0}^{n_o}v_{i,j,k,l,m} T_{o, j}T_{o, k}T_{o, l}T_{o, m}$ & $T^4_{o, 0}+\sum_{j,k,l,m=1}^{n_o}v_{0,j,k,l,m} T_{o, j}T_{o, k}T_{o, l}T_{o, m}$\\ 
        
    \end{tabular}
    \vspace*{0.3cm}

    The second difference between the model runs is that the T4 runs show a wider region of multistability in the interval $C\in[0.012\si{\kilogram\metre^{-2}\second^{-1}}, 0.0127\si{\kilogram\metre^{-2}\second^{-1}}]$. This is again assumed to be a result of higher order terms interacting in the T4 model equations, when compared to the DE runs, where the linearisation removes the higher order terms.
    
    To investigate the properties of the attractors in the regions of multiple stability, the attractors were projected onto the ocean variables $(\psi_{o, 1}, T_{o, 1})$ and $(\psi_{o, 2}, T_{o, 2})$ to visualise the behaviour. Looking at these projections we can see that there are three qualitatively distinct attractor behaviours, which correspond to three distinct flow behaviours in the ocean. We project all three attractors onto the same figure for comparison, shown in Figure \ref{fig:temp_bif_diag_C}, with the orange attractor shown for $C=0.009\si{\kilogram\metre^{-2}\second^{-1}}$, and the other two attractors shown for $C=0.0125\si{\kilogram\metre^{-2}\second^{-1}}$. In the interval $C\in [0.008\si{\kilogram\metre^{-2}\second^{-1}}, 0.009\si{\kilogram\metre^{-2}\second^{-1}}]$, there exists two attractors, where the attractor shown in orange displays LFV with respect to the variables $(\psi_{o, 1}, T_{o, 1})$. The orange attractor becomes unstable when $C>0.009\si{\kilogram\metre^{-2}\second^{-1}}$. This multistability has not been found in other studies using the linear MAOOAM model. The LFV in the first ocean modes suggests that the variability in the ocean temperature and streamfunctions is prominently impacted by the single gyre oscillation in this case. 
    
    The pink attractor, which becomes stable for values $C>0.012\si{\kilogram\metre^{-2}\second^{-1}}$, presents LFV over the variables $(\psi_{o, 2}, T_{o, 2})$, signifying that the dynamics of the flow are impacted by the double gyre dynamics. The other attractor (blue) present does not show LFV with any of the modes and this signifies that flows associated with this attractor do not present the same LFV or the same ocean dynamics as the other two attractors. The blue attractor becomes unstable for values $C>0.01275\si{\kilogram\metre^{-2}\second^{-1}}$. This second bifurcation was found in~\citet{vannitsem2017} using the linear version of the model, however the region of multistability in this interval was not found using the linear model. We have conducted long model runs to ensure the stability of both attractors in this interval and found that both attractors remain stable for at least $3 \times 10^7$ model days. This region of multistability was found in both the T4 and DE model runs, however it was found that the region of multistability is extended in the T4 model, when compared with the DE model.

    Following the analysis of Vannitsem et al.~\citep{vannitsem2017}, we project the attractors onto the variables $(\psi_{\mathrm{a}, 1}, \psi_{o, 1}, T_{o, 1})$ and $(\psi_{\mathrm{a}, 1}, \psi_{o, 2}, T_{o, 2})$ to visualise the degree of ocean-atmosphere coupling in the attractors. In Figure \ref{fig:3d_ocn_atm_C} we show all three attractors, for the same values of $C$ as before. In each image there is one attractor that stabilises around an unstable orbit that varies across all three variables over a decadal time scale. The image displaying $(\psi_{\mathrm{a}, 1}, \psi_{o, 2}, T_{o, 2})$ shows that the attractor coloured in pink presents the same oscillating behaviour over the three variables as the attractor found in ~\citet{vannitsem2017}. In addition, we also recover the attractor found by Vannitsem that does not present LFV (blue). The novel result here is that we have found an additional attractor (orange), for lower values of $C$, that also displays LFV in the coupled ocean-atmosphere system, but across the ocean-atmosphere modes $(\psi_{\mathrm{a}, 1}, \psi_{o, 1}, T_{o, 1})$.

    \begin{figure}
        \centering
        \subfloat[LLE of the DE and T4 model runs, for varying $C$ and $\varepsilon=0.7$.]{\includegraphics[width=0.46\textwidth]{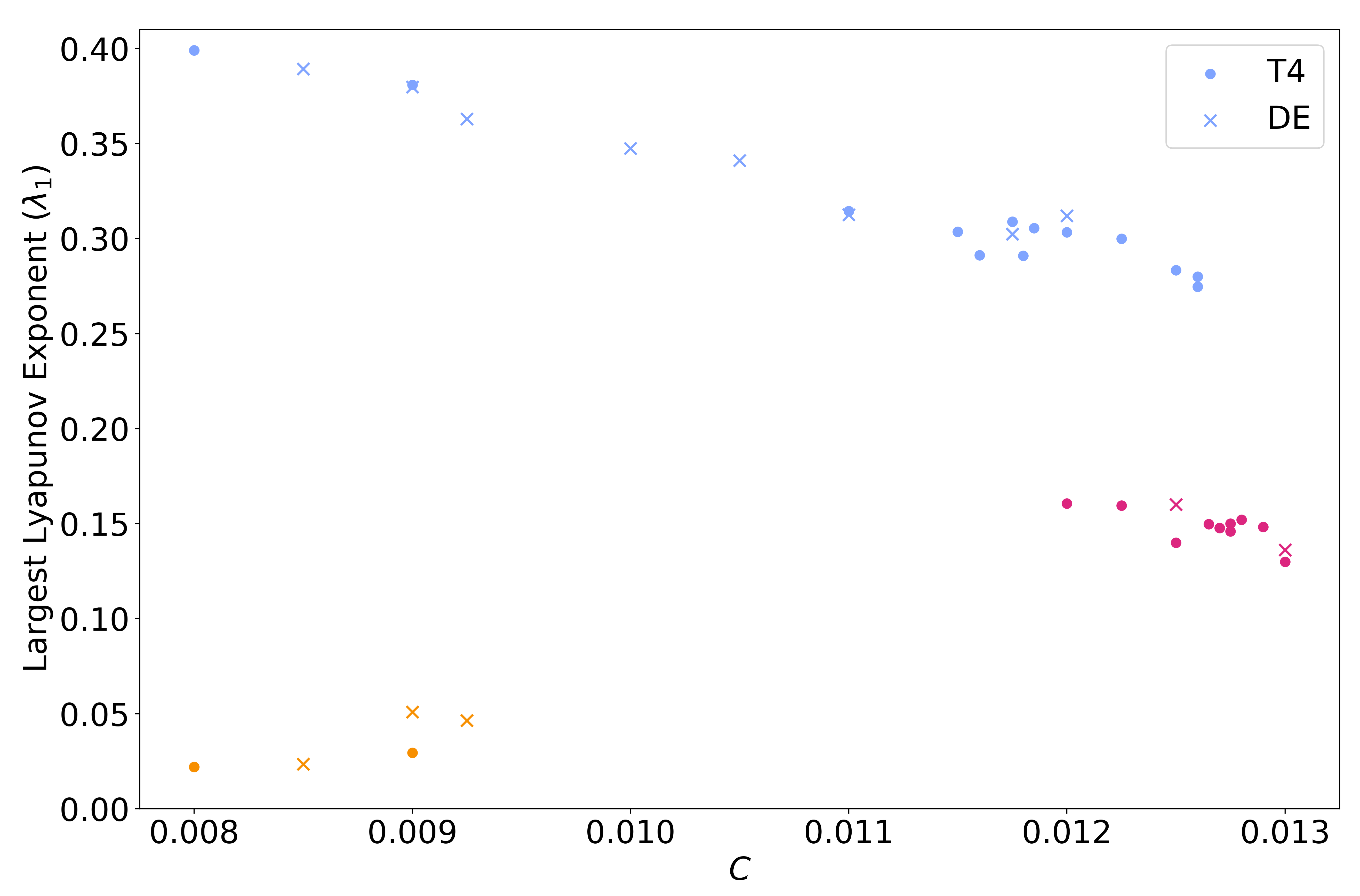}}
        \hspace{0.8cm}
        \subfloat[LLE of the DE and T4 model runs, for $C=0.01175\si{\kilogram\metre^{-2}\second^{-1}}$ and varying $\varepsilon$.]{\includegraphics[width=0.46\textwidth]{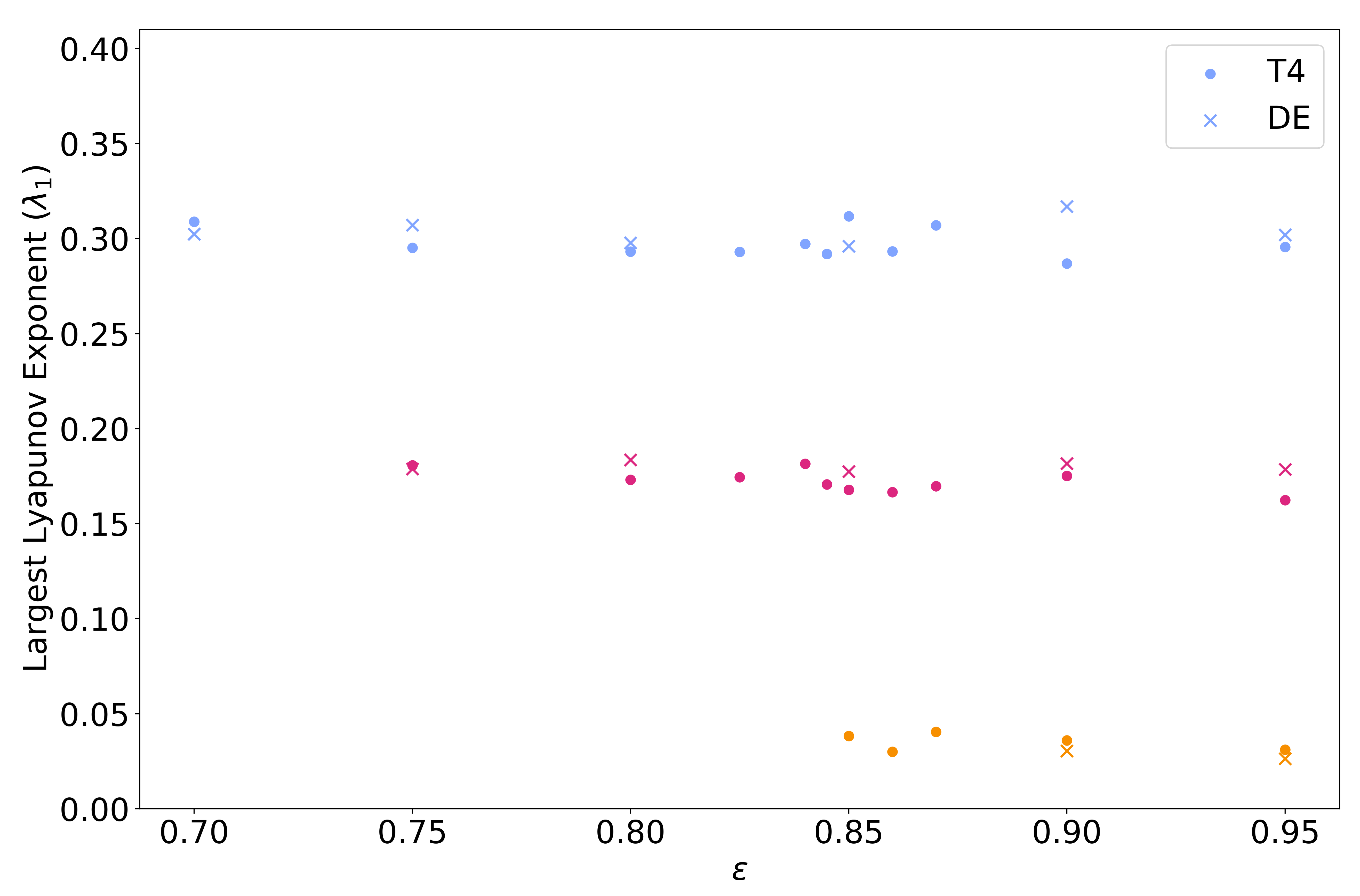}}
        \caption{The Largest Lyapunov Exponents (LLE), shown in days$^{-1}$, where the results are colour coded based on the attractor behaviour, as described in Sections \ref{subsec:temp_multistabilities_c} and \ref{subsec:temp_multistabilities_ep}.}
        \label{fig:LLE}
    \end{figure}

    To visualise the resulting flow in the ocean, given the attractor behaviour, we have created videos showing the ocean streamfunction and temperature profiles, given the location in the projected 2D state-space, which can be found in the supplementary materials. The videos are named based on the colour coding used in the above plots. The attractor coloured in blue shows a persistent positive value for the $T_{o, 2}$ variable, leading to a persistent double gyre temperature anomaly in the ocean temperature. In addition no clear LFV is present in the videos. The pink attractor shows oscillating behaviour over the variables $T_{o, 2}$ and $\psi_{o, 2}$ with timescales of approximately 70 model years, leading to transitions between a double and quadruple gyre temperature anomaly in the ocean. Similarly, a clear oscillation is seen in the ocean streamfunction where the most prominent variables oscillate between $\psi_{o, 2}$ and $\psi_{o, 6}$ over the same time period. Lastly, in the case of the orange attractor, no clear double gyre appears in the ocean temperature profile, and the LFV instead manifests in the first ocean mode $\phi_{o, 1}$. This leads to LFV with a timescale of 80-100 model years over the temperature variables $T_{o, 1}$, $T_{o, 5}$, and $T_{o, 7}$.

\subsubsection{Lyapunov Stability - \texorpdfstring{$C$}{}}\label{subsec:Lyp_c}
    To analyse the stability properties of the attractors that were found we calculated the Lyapunov properties of the stable attractors, focusing on values of $C$ identified in the previous section where multiple stable attractors exist. For more details on calculating Lyapunov properties in coupled ocean-atmosphere models see~\citet{vannitsem2016, vannitsem2017, vannitsem2019, vannitsem2020}.
    
    We present the largest Lyapunov exponents (LLE) in Figure \ref{fig:LLE} (a), as a function of $C$, where again there is a multistability present in the two intervals identified. We have used the same colour coding as in Section \ref{subsec:temp_multistabilities_c} to display which Lyapunov exponent is associated with which attractor. Approximately, the midlatitudes have a synoptic forecast time scale of less than two week~\citep{Lorenz_1982}, with larger scale process having a larger forecasting time~\citep{Lorenz_1969}. At the synoptic scale, this would correspond to LLE of approximately 0.2-0.3 days$^{-1}$. In the MAOOAM model there is a general decrease in the magnitude of the LLEs as $C$ increases due to the increase in coupling between the ocean and atmosphere, resulting in the flow instabilities from the atmospheric dynamics being reduced~\citep{vannitsem2017}. However, it is clear that the orange branch displays significantly smaller LLEs than the other two branches, for relatively small values of $C$. Other studies have shown that for low values of $C$ the magnitude of LLEs do decrease~\citep{vannitsem2017}, however this was observed for values $C\le 0.0015\si{\kilogram\metre^{-2}\second^{-1}}$, and we did not carry out runs for values of $C$ this low due to these ocean-atmospheric coupling values being unrealistic in the real earth system.

    \begin{figure}
        \centering
        \subfloat[Lyapunov spectra where the orange attractor is shown for $C=0.009\si{\kilogram\metre^{-2}\second^{-1}}$, and the other two for $C=0.0125\si{\kilogram\metre^{-2}\second^{-1}}$, where $\varepsilon=0.7$.]{\includegraphics[width=0.46\textwidth]{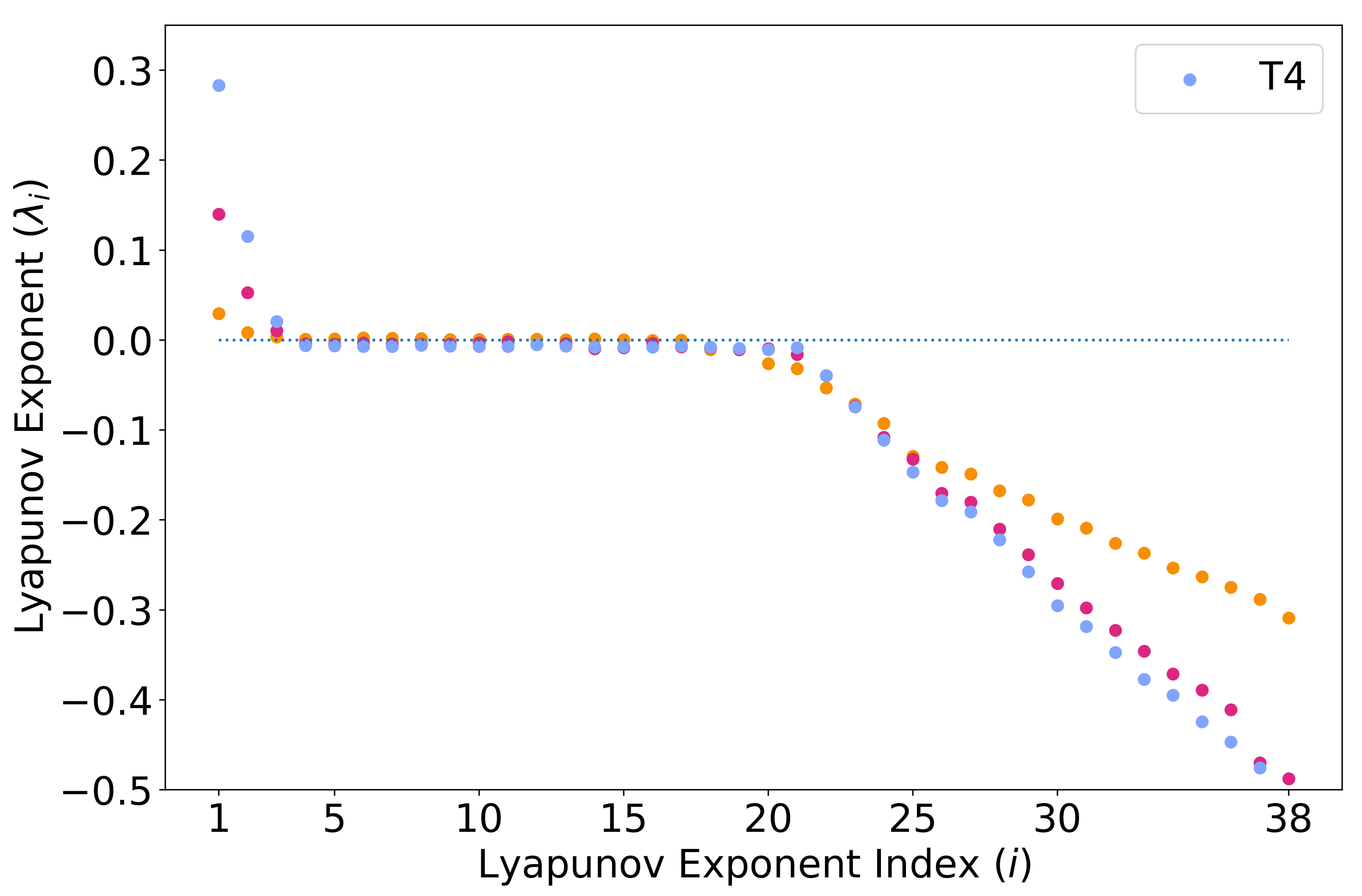}}
        \hspace{0.8cm}
        \subfloat[Lyapunov spectra for $\varepsilon=0.85$ and $C=0.01175\si{\kilogram\metre^{-2}\second^{-1}}$]{\includegraphics[width=0.46\textwidth]{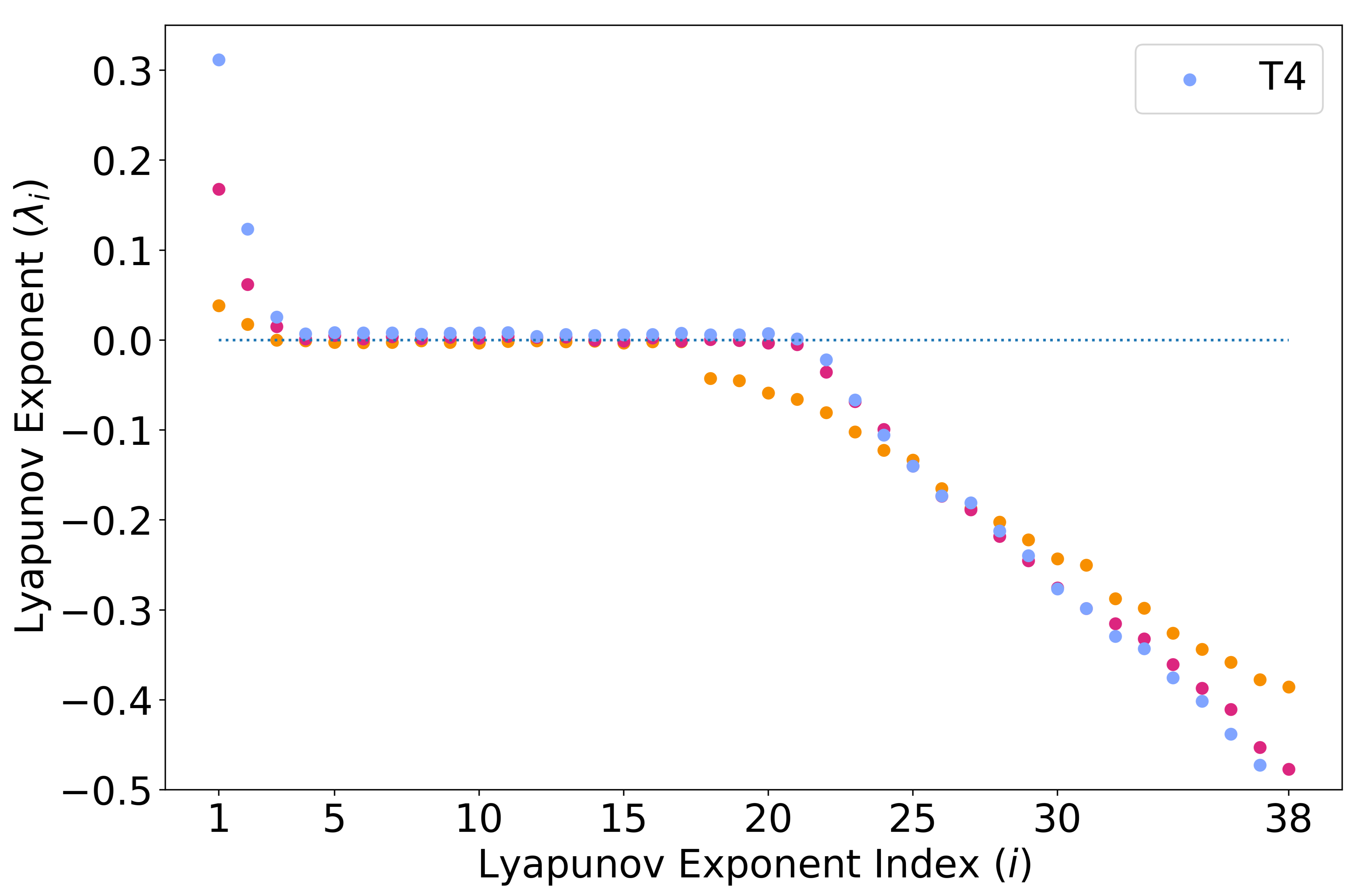}}
        \caption{The Lyapunov spectra, shown in days$^{-1}$, for the three distinct attractors found are plotted to compare the magnitude of the largest Lyapunov exponent $\lambda_1$, and the number of near zero Lyapunov exponents.}
        \label{fig:LSectrum}
    \end{figure}

    \begin{figure}
        \centering
        \includegraphics[width=\textwidth]{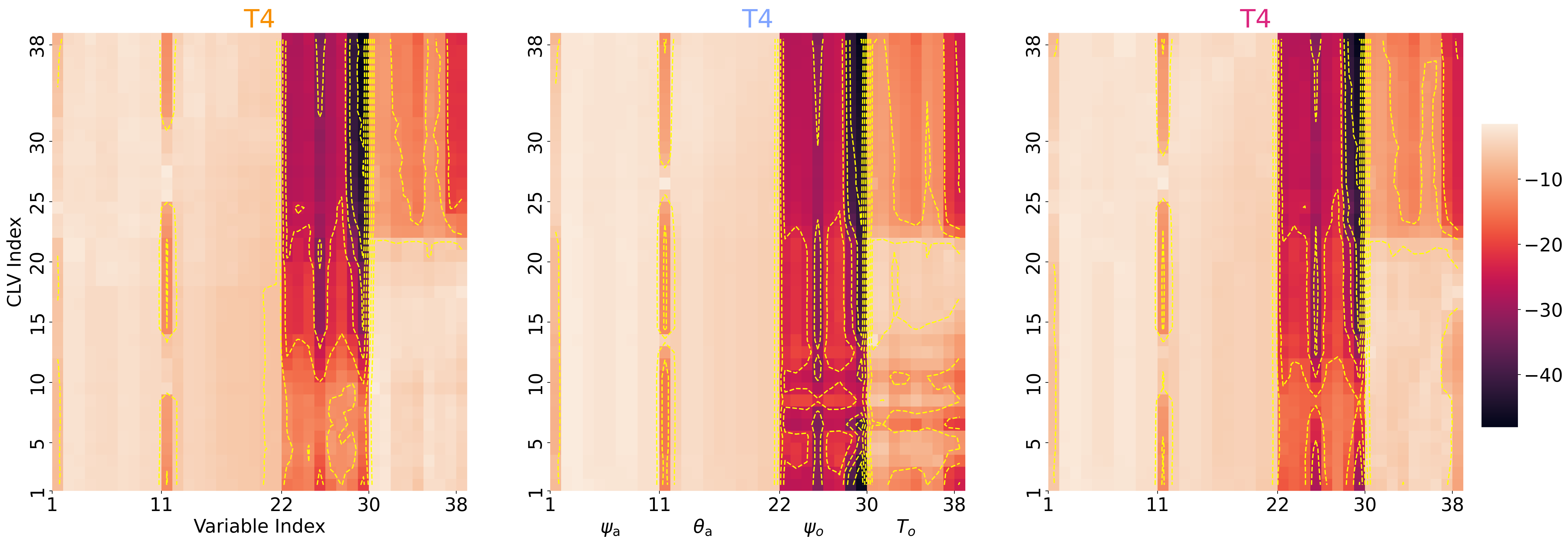}
        \caption{The variance of the CLVs (shown on a $\log_{10}$ scale), for each of the three distinct attractors shown in Figure \ref{fig:temp_bif_diag_C}. The attractors are designated by the colour of their title, which corresponds to the previous sections. The orange attractor CLVs are calculated for $\varepsilon=0.7$, and $C=0.009\si{\kilogram\metre^{-2}\second^{-1}}$, and the other two attractors' CLVs are calculated with $\varepsilon=0.7$ and $C=0.0125\si{\kilogram\metre^{-2}\second^{-1}}$.}
        \label{fig:CLV_c}
    \end{figure}

    In Figure \ref{fig:LSectrum} (a) we present the Lyapunov spectra of the three distinct attractors identified while varying the level of ocean-atmosphere coupling. The figure shows that the two attractors at $C=0.0125\si{\kilogram\metre^{-2}\second^{-1}}$ (blue and pink), both have 18 near zero Lyapunov exponents, 17 negative exponents, and three positive exponents. However the positive Lyapunov exponents are approximately double in one of the attractors compared with the other, showing that the rate of divergence of initial conditions will be much greater in the blue attractor. The novel attractor found in this study (orange), presents significantly smaller magnitude positive Lyapunov exponents than the other attractors, and has a lower number of near zero Lyapunov exponents, implying that the attractor exists on a lower dimensional chaotic manifold than the other attractors.
    The smaller positive magnitude Lyapunov exponents for two of the three attractors occurs due to these attractors existing around unstable periodic orbits that produce the LFV on decadal timescales and therefore increases the predictability. 
    
    The extent to which coupling between the ocean and atmosphere is responsible for the LFV can be visualised using the variance of Covariant Lyapunov Vectors (CLVs)~\citep{vannitsem2016}. The CLVs provide a covariant basis of the tangent linear space of the system. In other words, the CLVs are vectors that form a basis and remain covariant with the flow, unlike forwards or backwards Lyapunov vectors~\citep{kuptsov2012}. Each covariant Lyapunov vector is stretched by the system dynamics by the corresponding local Lyapunov exponent. Each CLV is made up of 38 components, one for every variable of the system. The CLVs were calculated at each time step, and we took the variance of each of the 38 vector components across time, for each of the 38 vectors. The variance measures the variability of the CLV component in the direction of a given variable. Variables from the atmosphere and ocean that both have high variance for the same CLV index implies that these variables are interacting, or influencing each other. Therefore variables that are coupled with one another will present higher variance for the same CLV index (horizontal rows on the diagram). This allows us to visualise which variables are coupled and have a greater impact on the system dynamics. In Figure \ref{fig:CLV_c} we present a heatmap of the variance ($\log_{10}$ scale) for the three distinct attractors. These plots show the CLV index on the y-axis, and the model variables on the x-axis. The variables are ordered as: 

    \begin{itemize}
        \item Atmospheric barotropic streamfunctions $\psi_{\mathrm{a}}$ (Index: 1-10)
        \item Atmospheric baroclinic streamfunctions $\theta_{\mathrm{a}}$ (Index: 11-21)
        \item Ocean barotropic streamfunctions $\psi_{\mathrm{o}}$ (Index: 22-29)
        \item Ocean temperature $T_o$ (Index: 30-38)
    \end{itemize}
    
    As expected, the majority of the variance is seen in the atmosphere as these variables are the components that contribute to the fast timescale dynamics of the system. All three attractors present coupling for CLV indices 13-22, where the ocean temperature variables present similar variability to the atmosphere variables. The near zero Lyapunov exponents (CLV index 4-22) in general have a larger projection on the ocean variables (columns 22-38). All three attractors present similar dynamics for the indicies greater than 22, where the ocean components have low projection on the dynamics, implying that the large magnitude negative Lyapunov exponents are predominantly caused by the atmosphere dynamics. However, the attractors that present LFV have higher variance for the indices 1-12, implying that in these attractors the ocean temperature is having a stabilising impact on the atmosphere dynamics, and that there is a greater level of coupling between the ocean and atmosphere. In addition, the orange attractor (heat map on the left hand side) shows greater coupling between all four components of the model, as the ocean streamfunctions show higher variance for the indices 1-10. This explains the low magnitude of the positive Lyapunov exponents for this attractor as the unstable atmosphere components present high levels of coupling with the stable ocean.

\subsection{Emissivity}\label{subsec:results_emissivity}
    To simulate the impact of climate change on the ocean-atmosphere dynamics in the MAOOAM model, the emissivity $\varepsilon$ is increased. Rising the emissivity acts as a proxy for rising levels of greenhouse gases, causing the atmosphere to `trap' a larger proportion of the outgoing longwave radiation, thus increasing the average ocean and atmosphere temperatures. We picked a single value of ocean-atmosphere coupling $C=0.01175\si{\kilogram\metre^{-2}\second^{-1}}$ to investigate. In the previous section this value of $C$ resulted in a single stable attractor, for $\varepsilon=0.7$.

\subsubsection{Multistabilities in Temperature - \texorpdfstring{$\varepsilon$}{}}\label{subsec:temp_multistabilities_ep}
    
    \begin{figure}
        \centering
        \includegraphics[width=\textwidth]{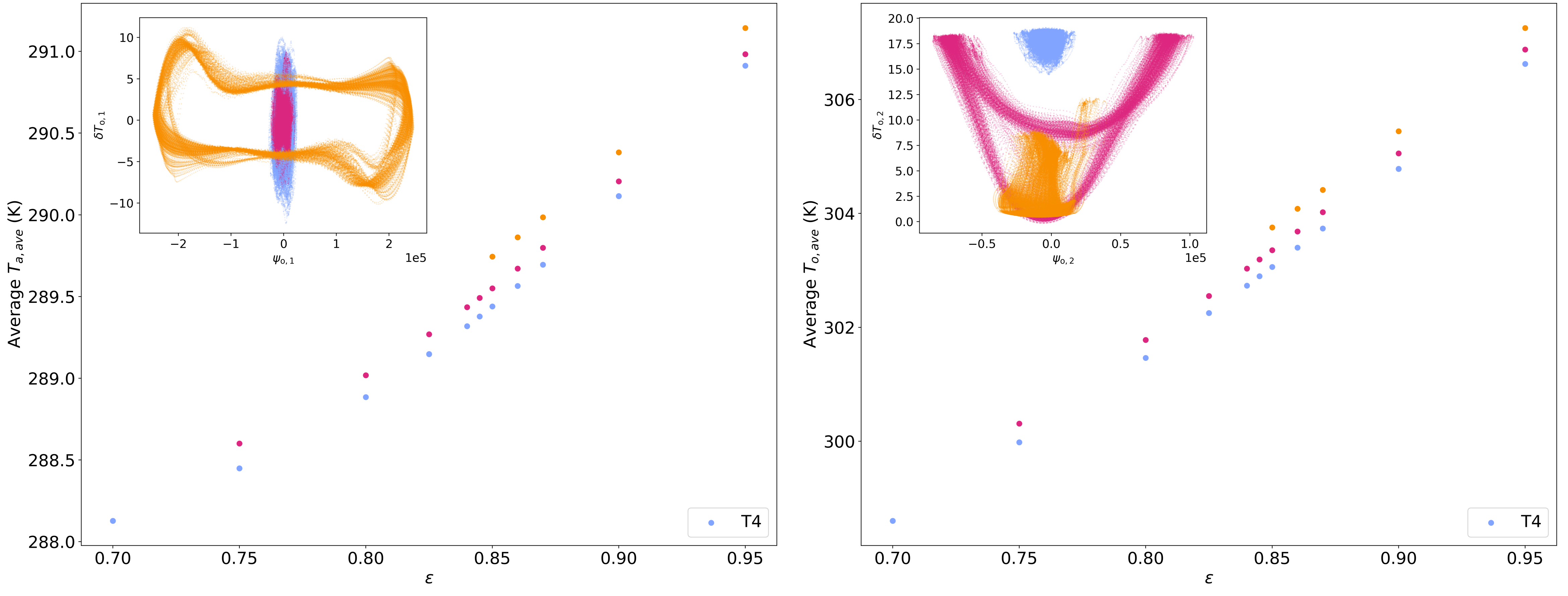}
        \caption{Temperature-$\varepsilon$ diagrams showing the averaged atmosphere (left) and ocean (right) temperatures. The different colours represent stable branches that present qualitatively different attractors. Here we only present the T4 model run results as the DE results produce the same results, and the temperature differences between the three branches of the DE runs are too small to be visible on these graphs. As in Figure \ref{fig:temp_bif_diag_C}, the attractors are projected onto the plane of ocean variables with the single gyre (left) and double gyre (right) for $\varepsilon=0.9$.}
        \label{fig:temp_bif_diag_ep}
    \end{figure}

    By using Temperature-$\varepsilon$ diagrams, shown in Figure \ref{fig:temp_bif_diag_ep}, that present the average spatial and temporal temperatures of stable attractors found numerically, we can see that as the emissivity increases bifurcations occur at two values, leading to an additional two stable branches. We have coloured the stable branches to display the attractors that present qualitatively distinct behaviour. On these images we only present the results from the T4 model runs as each of the T4 and DE model runs presented the similar trajectory behaviour, but the resulting average temperature of the three distinct branches in the DE runs were too close to be visible on the graph. This is because of the higher order non-linear terms being removed in the linearised version, as explained in Section \ref{subsec:temp_multistabilities_c}.
    
    We have taken the three stable attractors that we found at $\varepsilon=0.9$ and projected these onto the planes $(\psi_{o, 1}, T_{o, 1})$ and $(\psi_{o, 2}, T_{o, 2})$, and these are also shown in Figure \ref{fig:temp_bif_diag_ep}. We can see that two of the attractors (shown in pink and blue) present the same behaviour as seen in Section \ref{subsec:results_coupling}, in addition, the highest temperature branch (shown in orange) qualitatively appears to have similarities with the attractor that showed a periodic behaviour with respect to $(\psi_{o, 1}, T_{o, 1})$ that we also saw in the previous section. Therefore, increasing the value of $\varepsilon$ appears to have the impact of providing stability to unstable branches. Figure \ref{fig:temp_bif_diag_ep} shows that the average temperatures in the atmosphere and ocean rise quickly as $\varepsilon$ is increased, but in addition there are multistabilies that are separated by up to $1^\circ K$ for the same emissivity values.

    \begin{figure}
        \centering
        \includegraphics[width=\textwidth]{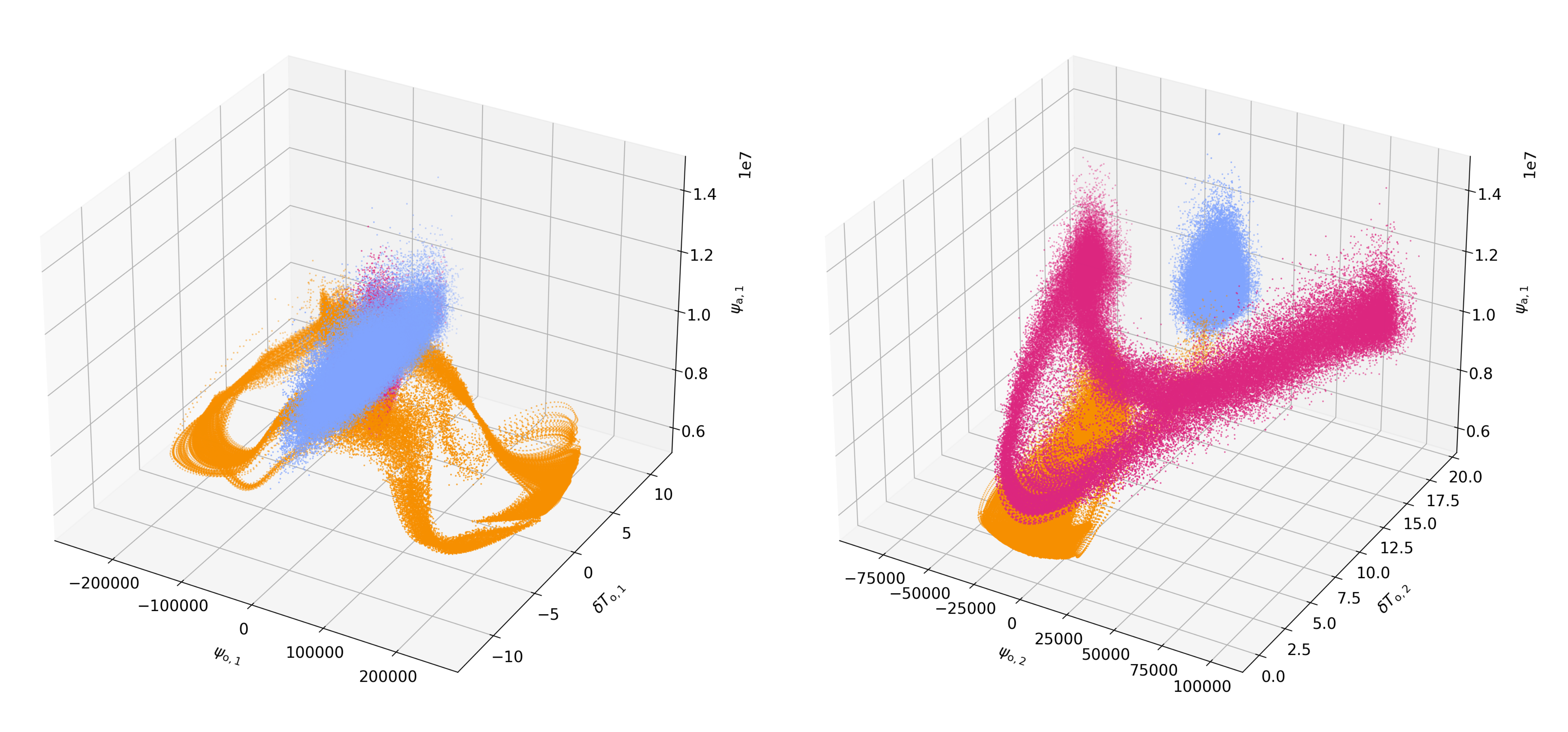}
        \caption{The three distinct attractors found at $C=0.01175\si{\kilogram\metre^{-2}\second^{-1}}$ and $\varepsilon=0.9$ are projected onto $(\psi_{\mathrm{a}, 1}, \psi_{o, 1}, T_{o, 1})$ and $(\psi_{\mathrm{a}, 1}, \psi_{o, 2}, T_{o, 2})$, left and right respectively. The attractors are coloured to match the diagrams shown in Figure \ref{fig:temp_bif_diag_ep}.}
        \label{}
    \end{figure}

    As in Section \ref{subsec:results_coupling}, we visualise the level of ocean-atmosphere coupling by projecting the three stable attractors onto the first atmosphere mode, and the single and double gyre ocean modes. As with this previous section, we see that the orange attractor displays an oscillatory behaviour, coupling the barotropic streamfunctions and the single gyre variables, and the pink attractor displays LFV with the double gyre variables. Again, we see two distinct flows where there exists coupling between the ocean and atmosphere, over decadal time periods. These results show that in the MAOOAM model, rising emissivity leads to multistabilies in the ocean-atmosphere system, that were not present at low levels of emissivity. From our model runs we have not found examples of trajectories switching between the stable branches, however all of our model runs were undertaken using a constant solar forcing. To understand the robustness of the attractors to forcing, further model runs will have to be undertaken.

    Similar to Section \ref{subsec:temp_multistabilities_c}, we have produced videos to show the resulting ocean streamfunction and temperature behaviour given the attractors. These videos can be found in supplementary materials. The qualitative behaviour of each of the attractors identified in this section is similar to that of the corresponding attractors (those sharing the same colours) in Section \ref{subsec:temp_multistabilities_c}. One minor difference in the results between the ocean-atmosphere coupling model runs and the emissivity model runs is that the LFV in the orange attractor over the first ocean mode $\phi_{o, 1}$ is more clearly defined. This is shown by the projection of the attractor on the plane $(\psi_{o, 1},\ T_{o, 1})$, where the oscillating behaviour over these variables contains less noise and variation. This change in the orange attractor could be caused by the increase in ocean-atmosphere coupling between the two runs, where we used the value of $C=0.009\si{\kilogram\metre^{-2}\second^{-1}}$ in Section \ref{subsec:temp_multistabilities_c}, and $C=0.01175\si{\kilogram\metre^{-2}\second^{-1}}$ in this section.
    
\subsubsection{Lyapunov Stability - \texorpdfstring{$\varepsilon$}{}}
    Following the format of Section \ref{subsec:Lyp_c}, we present the LLEs in Figure \ref{fig:LLE} (b), where $\varepsilon$ is varied, and we fix the ocean-atmosphere coupling at $C=0.01175\si{\kilogram\metre^{-2}\second^{-1}}$. As the value of $\varepsilon$ is increased we see additional stable attractors appear, however the value of the LLEs on each stable branch does not alter with emissivity. This is because the emissivity has the impact of increasing the temperature of the layers evenly in space. The atmospheric layers in the model are driven from the meridional gradient in solar insolation, leading to baroclinic instability. In the current model set up the emissivity has no impact on this temperature gradient. To more closely model the expected outcomes of global heating, model runs should be undertaken where the rising emissivity reduces the temperature gradient between the Arctic and the equator~\citep{francis2012, rantanen2022}.

    We have presented the Lyapunov spectra of the three attractors at $\varepsilon=0.9$, for $C=0.01175\si{\kilogram\metre^{-2}\second^{-1}}$ in Figure \ref{fig:LSectrum} (b). Interestingly, the orange attractor, with the lowest magnitude LLE, shows a significantly lower number of near zero Lyapunov exponents, compared to the other two attractors and also the orange attractor presented in Section \ref{subsec:Lyp_c}. There is a clear drop in the magnitude of the Lyapunov exponents at the index 18. This is an interesting result as it implies that the Lyapunov dimension of this attractor is significantly lower than the other attractors. This difference between the Lyapunov spectra of the orange attractors in Figure \ref{fig:LSectrum} (a) and (b) could be caused by the increase in ocean-atmosphere coupling between the two model runs, from $C=0.009\si{\kilogram\metre^{-2}\second^{-1}}$ to $C=0.0175\si{\kilogram\metre^{-2}\second^{-1}}$ in this section.

\section{Discussion}\label{sec:discussion}
    We have described the novel multistabilities found in a reduced order atmosphere-ocean model, resulting from not linearising the longwave radiation terms. The modifications made to the MAOOAM model have resulted in several features that were not present in the original linearised version with fixed reference temperature. This study introduced two new versions of the model (the dynamic equilibria and non-linear versions), and compared the results of these versions with the existing linear model. The properties of the new attractor found, as well as the region of multistability, were analysed qualitatively and by using the Lyapunov properties of the attractors.
    
    We have demonstrated in the reduced order ocean-atmosphere model, MAOOAM, that by modifying the linearisation of the longwave radiation terms we can obtain three qualitatively distinct stable attractors, with intervals of multistability for certain parameters. Interestingly, the dynamic equilibria version of the model, which includes the same linearisation as the original model but allows the zeroth order equilibria temperature to change with time, presents similar dynamics as the fully non-linear version for most parameter values. All three distinct attractors can be obtained in the dynamic equilibria (DE) version, where there are two attractors that present LFV while representing largely different coupled ocean-atmosphere flows. In addition, the DE version of the model can be run in the same length of time as the fixed reference temperature version, which is almost an order of magnitude faster than the non-linear version of the model.

    Two of the distinct attractors present low frequency variability (LFV) behaviour. One of the attractors, which displays LFV over the second ocean mode $\phi_2$, has been identified in the linear model. In this study we found an additional attractor that displays LFV over the first ocean mode $\phi_1$, which has not been identified in the linear model. This attractor has a longer time scale ($\sim 80-100$ years) compared with the first attractor (which displays a timescale of approximately $70$ years). In addition the two attractors display marked differences in the ocean and atmosphere flows. With one attractor producing the double gyre behaviour, similar to what is observed over the North Atlantic, and the other attractor displaying a more complex flow, where the main relationships are with the first ocean mode $\phi_1$ and the fifth ocean mode $\phi_5$. This attractor displayed the greatest degree of coupling between the ocean and atmosphere, comparing with the other stable attractors, where all four variables are coupled. This leads to significantly lower positive Lyapunov exponents, which implies that this attractor would have a longer forecasting window. Further studies will need to be done to find if this attractor describes a real-world ocean-atmosphere flow. 

    A key reason for undertaking this study, and not linearising the longwave radiation terms, was to investigate the potential of tipping between multistabilities. While we found cases of distinct attractors for the same parameter values, we could not produce trajectories that switched intermittently between the stable branches, though it is not possible to rule out the possibility of such trajectories existing. However, in the current model setup all external forcings are stable with respect to time. To test the robustness of the stability of each attractor model runs could be undertaken, where stochastic forcing or perturbations are included, to see if there is the potential for noise induced tipping between the stable branches. Another potential source of tipping could come from periodic cycles, such as the annual solar cycle as implemented in a similar linearised model~\citep{vannitsem2017}.

    With rising global temperatures, investigating the possibility and impact of tipping points in the climate is of great importance. Understanding how rising temperatures could impact existing multi-decadal patterns in the climate could lead to a better understanding of how established climate patterns may look in the future. Or if there is the possibility of abrupt transitions from one regime to another. This paper has introduced a modified model that produces both multistabilities, as well as attractors that present LFV, that become stable for rising emissivity. These properties are of interest as it facilitates the study of attractors that allow forecasting well beyond the atmospheric Lyapunov time, as well as the potential of tipping from one attractor to the other. These aspects will be investigated in the future.

\section*{code availability}
The code used to obtain the results is a new version (v0.2.6) of qgs~\citep*{demaeyer2020} that was recently released on GitHub: \url{https://github.com/Climdyn/qgs} and Zenodo~\citep*{qgs_model}.

\section*{acknowledgements}
This project has received funding from the European Union’s Horizon 2020 research and innovation programme under the Marie Sklodowska-Curie grant agreement No.956170. MC was funded as Research Director with the Belgian National Fund of Scientific Research.

\section*{conflict of interest}
The authors declare no conflict of interest.

\section*{supporting information}

The documentation manual of the new qgs version associated with the new temperature scheme is provided as a supplementary material to this article.

The videos discussed in Section \ref{subsec:temp_multistabilities_c} and \ref{subsec:temp_multistabilities_ep} can be accessed in the below table.

\begin{tabular}{ |p{4cm}||p{3cm}|p{2cm}|p{2cm}|  }
    \hline
    \multicolumn{4}{|c|}{Video list} \\
    \hline
    Link& $C$ value & $\varepsilon$ value & Attractor colour\\
    \hline
    \href{https://doi.org/10.5446/60104}{https://doi.org/10.5446/60104} & $C=0.009\si{\kilogram\metre^{-2}\second^{-1}}$ & $\varepsilon=0.7$ & Orange\\
    \href{https://doi.org/10.5446/60105}{https://doi.org/10.5446/60105} & $C=0.0125\si{\kilogram\metre^{-2}\second^{-1}}$ & $\varepsilon=0.7$ & Blue\\
    \href{https://doi.org/10.5446/60106}{https://doi.org/10.5446/60106} & $C=0.0125\si{\kilogram\metre^{-2}\second^{-1}}$ & $\varepsilon=0.7$ & Pink\\
    \href{https://doi.org/10.5446/60107}{https://doi.org/10.5446/60107} & $C=0.01175\si{\kilogram\metre^{-2}\second^{-1}}$ & $\varepsilon=0.9$ & Blue\\
    \href{https://doi.org/10.5446/60108}{https://doi.org/10.5446/60108} & $C=0.01175\si{\kilogram\metre^{-2}\second^{-1}}$ & $\varepsilon=0.9$ & Orange\\
    \href{https://doi.org/10.5446/60109}{https://doi.org/10.5446/60109} & $C=0.01175\si{\kilogram\metre^{-2}\second^{-1}}$ & $\varepsilon=0.9$ & Pink\\
    \hline
\end{tabular}

\bibliography{MyBib}

\graphicalabstract{Figures/graphicalabstract}{The climate system contains numerous non-linear interactions that produce chaotic behaviour and provides the possibility of multiple stationary solutions as well as tipping between solutions. This study investigates the impact of implementing the non-linear Stefan-Bolzmann law in a reduced order model. This model produces multiple stationary oscillating solutions. These solutions are a potential method of extending forecasts of the weather, which is limited due to sensitivity to initial conditions.}

\end{document}